\documentclass[twocolumn,letterpaper,pdftex]{article}

\usepackage{graphicx}
\usepackage[fleqn]{amsmath}
\usepackage{newtxtext}
\usepackage[varg]{newtxmath}
\usepackage{url}
\usepackage{multirow}
\usepackage{amsmath}
\usepackage{cite}
\usepackage[top=2cm, bottom=2cm, left=1.5cm, right=1.5cm]{geometry}

\usepackage{hyperref}

\title{Performance comparison of the two reconstruction methods for stabilizer-based quantum secret sharing}
\author{Shogo Chiwaki$^{{\dagger} {\rm a)}}$, Ryutaroh Matsumoto$^{{\dagger}, {\dagger} {\dagger} {\rm b)}}$}

\makeatletter
\renewcommand{\section}{\@startsection{section}{1}{\z@}%
   {-1\baselineskip \@plus-1mm \@minus-.5mm}%
   {1\baselineskip \@plus1mm \@minus.5mm}%
   {\reset@font\normalsize\bfseries\mathversion{bold}}}
\makeatother

\makeatletter
\renewcommand{\subsection}{\@startsection{subsection}{2}{\z@}%
   {-1\baselineskip \@plus-1mm \@minus-.5mm}%
   {1\baselineskip \@plus1mm \@minus.5mm}%
   {\reset@font\normalsize}}
\makeatother

\begin{document}
\maketitle

\begin{abstract}
Stabilizer-based quantum secret sharing has two methods to reconstruct a quantum secret: The erasure correcting procedure and the unitary procedure.
It is known that the unitary procedure has a smaller circuit width. On the other hand, it is unknown which method has smaller depth and fewer circuit gates.
In this paper, it is shown that the unitary procedure has smaller depth and fewer circuit gates when the circuits are designed for quantum secret sharing using $[[5, 1, 3]]$ binary stabilizer codes.
\end{abstract}

\renewcommand{\thefootnote}{\fnsymbol{footnote}}
\footnote[0]{$^{{\dagger}}$The author is with Department of Information and Communications Engineering, Tokyo Institute of Technology, Tokyo, 152-8550, Japan.}
\footnote[0]{$^{{\dagger}{\dagger}}$The author is with Department of Mathematical Sciences, Aalborg University, Aalborg, Denmark.}
\footnote[0]{a) chiwaki.s.aa@m.titech.ac.jp}
\footnote[0]{b) ryutaroh@ict.e.titech.ac.jp}

\section{Introduction}
Quantum secret sharing \cite{quantum secret sharing} is a cryptographic scheme to encode a quantum secret into multiple pieces of quantum information (called shares) and distribute shares to participants so that qualified sets of participants can reconstruct the secret but forbidden sets can gain no information about the secret.
Quantum secret sharing protects the quantum secret, in the same way that classical secret sharing protects the classical secret.

A stabilizer code \cite{stabilizer code} is a class of quantum codes which can detect and correct errors.
$[[n, k, d]]$ stabilizer codes are encoding $k$ qudits into $n$ qudits, detecting $d-1$ quantum errors, and correcting $d-1$ quantum ``erasures'' \cite{erasure}. Here, an erasure means an error whose position is known.

Quantum secret sharing can be constructed by stabilizer codes \cite{quantum secret sharing}, \cite{erasure to share}.
Any particular sequence of $n$ qudits which the encoder might transmit is called a codeword.
Each qudit of the codeword of a stabilizer code is a share of the quantum secret sharing. 
If an error-correcting code can correct $d-1$ erasures, any $n-(d-1)$ shares can recover the initial state by treating the missing $d-1$ shares as erasures.

Quantum secret sharing using stabilizer codes has two different procedures to reconstruct a secret from a set of shares.
One is an erasure correcting procedure \cite{quantum secret sharing}, and the other is a unitary procedure \cite{unitary procedure}.
Details of both procedures are described in the next section.

Depth, width, and the number of circuit gates are performance indices of quantum circuits.
Depth is the number of steps needed to complete the circuit operation.
Width is the number of qudits required to run the circuit.
Circuit gates consist of measurements and unitary gates acting on one or two qudits in a quantum circuit.

In terms of width and the number of measurements, the unitary procedure is better than the erasure correction procedure \cite{unitary procedure}.
However, we do not know which is better in terms of depth and the number of unitary gates.

In this paper, a quantum circuit for each method is designed for quantum secret sharing using $[[5, 1, 3]]$ binary stabilizer codes.
We compare the erasure correcting procedure with the unitary procedure by counting depth and the number of circuit gates of both circuits.	
We find that depth and the number of circuit gates of the unitary procedure are smaller than those of the error correcting procedure.

\section{Two procedures to reconstruct a quantum secret}
There are two procedures to reconstruct a quantum secret.
One is an erasure correcting procedure \cite{quantum secret sharing}, the other is a unitary procedure \cite{unitary procedure}. Detail of both procedures is as follows.

\subsection{Erasure correcting procedure}
Participants put shares in the corresponding places of the codeword of the stabilizer code.
The participants add new qudits to the empty places and regards them as erasures.
If the erasures are correctable, the participants get the original codeword by performing erasure correction.
Decoding the codeword reveals the secret \cite{quantum secret sharing}.

\subsection{Unitary procedure}
There are multiple sets of shares that can reconstruct a secret.
Each set has its own unitary transformation to reconstruct the secret.
When participants get a set of shares that can reconstruct a secret, the secret can be found by performing the corresponding unitary transformation \cite{unitary procedure}.

\section{Condition of comparison and design method of circuit}
\subsection{Condition of comparison}
We reconstruct the secret from the third, fourth and fifth qubits of the codeword of the $[[5, 1, 3]]$ binary stabilizer codes.

We design and draw quantum circuits using Q\#, which is Microsoft's open-source programming language for developing and running quantum algorithms \cite{Qsharp}.
The program code that we have written can be found in the ipynb format as auxiliary files.
The auxiliary file erasure\_correcting\_procedure.ipynb contains the designing of the erasure correcting procedure and the visualisation of the circuit.
The auxiliary file unitary\_procedure.ipynb contains the designing of the unitary procedure and the visualisation of the circuit.
The circuit gates used in quantum circuits are measurements, CNOT gates, X gates and H gates.

We denote a quantum secret as $\alpha|0\rangle+\beta|1\rangle$ ($\alpha$, $\beta$ are complex number, and $|\alpha|^2+|\beta|^2=1$).
We encode the message into $\alpha|0_L\rangle+\beta|1_L\rangle$ according to ``EncodeIntoFiveQubitCode'' \cite{EncodeIntoFiveQubitCode} defined by Q\#, and $|0_L\rangle$,$|1_L\rangle$ are as follow.

\begin{equation*}
\begin{split}
|0_L\rangle
=-\frac{1}{4}\bigl\lbrack
&|11111\rangle+|01101\rangle+|10110\rangle+|01011\rangle\\
+&|10101\rangle-|00100\rangle-|11001\rangle-|00111\rangle\\
-&|00010\rangle-|11100\rangle-|00001\rangle-|10000\rangle\\
-&|01110\rangle-|10011\rangle-|01000\rangle+|11010\rangle\bigr\rbrack.\\
\end{split}
\end{equation*}

\begin{equation*}
\begin{split}
|1_L\rangle
=-\frac{1}{4}\bigl\lbrack
&|00000\rangle+|10010\rangle+|01001\rangle+|10100\rangle\\
+&|01010\rangle-|11011\rangle-|00110\rangle-|11000\rangle\\
-&|11101\rangle-|00011\rangle-|11110\rangle-|01111\rangle\\
+&|10001\rangle-|01100\rangle-|10111\rangle+|00101\rangle\bigr\rbrack.\\
\end{split}
\end{equation*}\\

\subsection{Design method of circuit of the erasure correcting procedure}
We used Q\# operations to design the circuit.
The operations we used are ``Recover'' \cite{Recover} and ``DecodeFromFiveQubitCode'' \cite{DecodeFromFiveQubitCode}.

Recover is an operation that gives an error-corrected codeword and takes three arguments.
The first argument is the quantum state to be corrected; Recover gives the corrected codeword by operating on this argument.
The second argument consists of encoding and decoding procedures.
The third argument is a two-dimensional array that describes the correcting operation according to the measurement results.

DecodeFromFiveQubitCode reconstructs the secret from the codeword of the $[[5, 1, 3]]$ binary stabilizer code.

We used the command ``\%trace'' \cite{trace} to visualize the designed circuit.
\%trace is only available in the Jupyter Notebook.
We redrew the circuit manually to a simpler version.

\subsection{Design method of circuit of the unitary procedure}
We designed the circuit by hand.

\section{Results}
\subsection{Designed circuits}
\begin{figure}[htbp]
  \begin{minipage}[b]{0.5\columnwidth}
    \centering
    \includegraphics[width=\columnwidth]{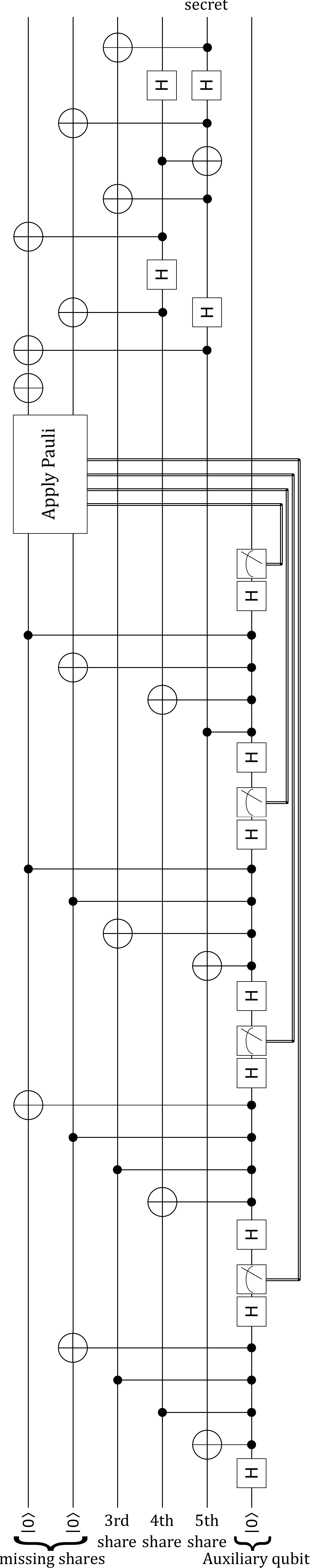}
    \caption{{$\scriptstyle \mbox{Circuit of the erasure}\atop \scriptstyle \mbox{correcting procedures}$}}
    \label{fig:erasure}
  \end{minipage}
  \begin{minipage}[b]{0.45\columnwidth}
    \centering
    \includegraphics[width=0.5\columnwidth]{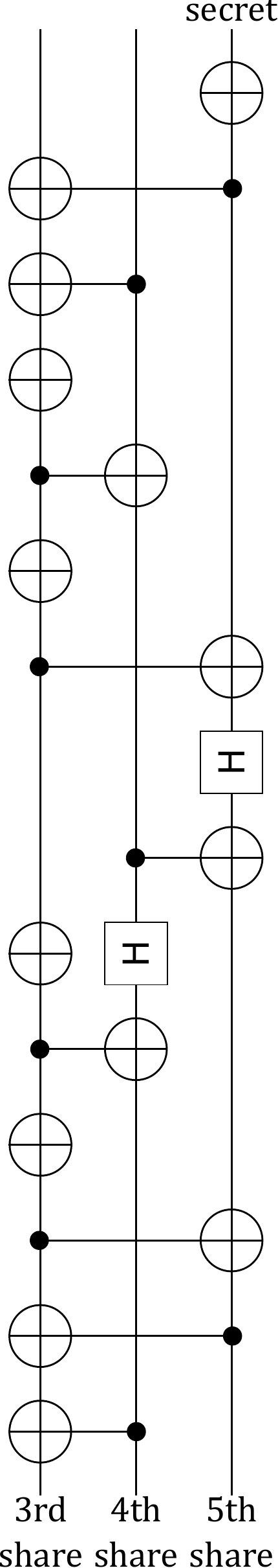}
    \caption{{$\scriptstyle \mbox{Circuit of}\atop \scriptstyle \mbox{unitary procedures}$}}
    \label{fig:unitary}
  \end{minipage}
\end{figure}

Fig.\ref{fig:erasure} is the quantum circuit of the erasure correcting procedure.
Fig.\ref{fig:unitary} is the quantum circuit of the unitary procedure.
The legend for the symbols in each figure is in Table \ref{tab:number}.

In a CNOT gate, the black dot is a control qubit and the $\bigoplus$ is a target qubit.
In a controlled-Z gate, one black dot is a control qubit and the other is a target qubit.
The controlled-Z gates do not change the result whichever black dot is used as the control qubit.
The controlled-Z gate consist of one CNOT gate and two H gates. In this paper, the controlled-Z gate is counted as one CNOT gate and two H gates.
The double line from the measurement is wire of classical bit.
The box marked as Apply Pauli in Fig.\ref{fig:erasure} is a unitary gate to correct errors whose content changes according to the four results of measurements.
The box marked as Apply Pauli has 0, 1 or 2 Pauli gates.

``3rd share'', ``4th share'' and ``5th share'' in both circuits are the third, fourth and fifth qubits of the codeword of the $[[5, 1, 3]]$ binary stabilizer codes respectively.
The $|0\rangle$s labeled missing shares are new qubits added to treat the missing shares as erasures.
The $|0\rangle$ written as Auxiliary qubit is a qubit to simplify the measurement.

\subsection{Depth, width, and the number of circuit gates}
We list depth, width and the number of circuit gates for both procedures in Table \ref{tab:number}.
We assume that the box marked Apply Pauli has no Pauli gates. This assumption is advantageous to the erasure correcting procedure.

\begin{table}[htb]
\caption{{$\scriptstyle \mbox{legend of each gate and}\atop \scriptstyle \mbox{depth, width, the number of circuit gates}$}}
\label{tab:number}
\centering
  \begin{tabular}{c|ccc}
    \hline
    &\multirow{2}{*}{legends}&unitary&erasure correcting\\
    &&procedures&procedures\\
    \hline \hline
    CNOT gates&\begin{minipage}{5mm}\centering\scalebox{0.5}{\includegraphics{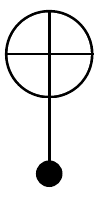}}\end{minipage}&9&23\\\hline
    X gates&\begin{minipage}{5mm}\centering\scalebox{0.5}{\includegraphics{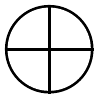}}\end{minipage}&5&1\\\hline
    H gates&\begin{minipage}{5mm}\centering\scalebox{0.5}{\includegraphics{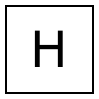}}\end{minipage}&2&28\\\hline
    measurements&\begin{minipage}{5mm}\centering\scalebox{0.5}{\includegraphics{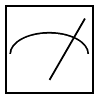}}\end{minipage}&0&4\\\hline
    controlled-Z gates&\begin{minipage}{5mm}\centering\scalebox{0.5}{\includegraphics{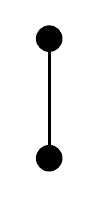}}\end{minipage}&&\\\hline
    depth&&15&38\\\hline
    width&&3&6\\\hline
  \end{tabular}
\end{table}

\section{Conclusion}
We count depth and the number of circuit unitary gates of the quantum circuits for both procedures in quantum secret sharing using $[[5, 1, 3]]$ binary stabilizer codes.
In this paper, the unitary procedure is smaller than the erasure correcting procedure in terms of depth and the total number of circuit unitary gates.
It has shown that the unitary procedure is better than the erasure correcting procedure in terms of the width and the number of measurements.
Therefore, we conclude that the unitary procedure is better than the erasure correcting procedure under the conditions of this paper.

We obtain an example of the relationship between both procedures.
In the future we will research whether the relationship between both procedures holds in the general case.



\end{document}